\documentstyle[epsfig,12pt]{article}  
\newcommand{\beq}{\begin{equation}}
\newcommand{\eeq}{\end{equation}}
\newcommand{\bea}{\vspace{0.25cm}\begin{eqnarray}}
\newcommand{\eea}{\end{eqnarray}}

\newcommand{\r}{\mbox{{\boldmath
$\rho$}}}

\newcommand{\qb}{\mbox{{\bf
q}}}

\def\lsim{\mathrel{\rlap{\lower4pt\hbox{\hskip1pt$\sim$}}
    \raise1pt\hbox{$<$}}}         
\def\gsim{\mathrel{\rlap{\lower4pt\hbox{\hskip1pt$\sim$}}
    \raise1pt\hbox{$>$}}}         
\begin{document}
\vspace*{-2cm}
 
\bigskip

\begin{center}

\renewcommand{\thefootnote}{\fnsymbol{footnote}}

  {\Large\bf
Radiative parton energy loss and 
jet quenching in high-energy heavy-ion collisions
\footnote{Supported by DFG-grant Schi 189/6-1}
\\
\vspace{.7cm}
  }
\renewcommand{\thefootnote}{\arabic{footnote}}
\medskip
{\large
  B.G.~Zakharov
  \bigskip
  \\
  }
{\it
Fakult\"at f\"ur Physik, Universit\"at Bielefeld\\
D-33501 Bielefeld, Germany\\
and\\
 L.D.~Landau Institute for Theoretical Physics,
        GSP-1, 117940,\\ Kosygina Str. 2, 117334 Moscow, Russia
\vspace{1.7cm}\\}

  {\bf
  Abstract}
\end{center}
{
\baselineskip=9pt
We study within the light-cone path integral approach \cite{Z1}
the effect of the induced gluon radiation 
on high-$p_{T}$ hadrons in high-energy heavy-ion collisions.
The induced gluon spectrum is represented in a new form which
is convenient for numerical simulations. For the first time,
computations are performed with a realistic parametrization
of the dipole cross section.
The results are in reasonable agreement with 
suppression of high-$p_{T}$ hadrons in $Au+Au$ collisions 
at $\sqrt{s}=200$ GeV observed at RHIC.
\vspace{.5cm}
\\
}

\noindent{\bf 1.} One of the most interesting results 
obtained at RHIC is the suppression of high-$p_{T}$ hadrons in 
$Au+Au$ collisions (for a review of the data, see \cite{RHIC_data}).
It is widely believed that parton energy loss due to 
the induced gluon radiation caused by multiple scattering in the 
quark-gluon
plasma (QGP) produced in the initial stage of nucleus-nucleus collisions
plays a major role in this phenomenon (usually called  
jet quenching) \cite{BDMPS,Z1,Z2,BDMS1,GLV1,W1} (for a review, 
see \cite{BSZ}).
The most general approach to the induced gluon emission
is the light-cone path integral (LCPI) approach developed in
\cite{Z1} (see also \cite{Z_YAF,Z3,BSZ}). 
It accurately treats  the mass and finite-size effects, and applies
at arbitrary strength of the Landau-Pomeranchuk-Migdal
(LPM) effect \cite{LP,Migdal}. Other available approaches
have limited domains of applicability, and can only be used either in 
the regime of strong (the BDMPS formalism \cite{BDMPS,BDMS1})
or weak (the GLV formalism \cite{GLV1}) LPM suppression (the GLV 
approach \cite{GLV1}, in addition, is restricted to the emission of soft
gluons). For this reason they can not be used for an 
accurate analysis of jet quenching
for RHIC (and LHC) conditions.

The LCPI approach expresses the gluon spectrum
through the solution of a two-dimensional Schr\"odinger
equation with an imaginary potential in the impact parameter plane.
The imaginary potential is 
proportional to the cross section of interaction
of the $\bar{q}qg$ system (for $q\rightarrow gq$ transition) with a 
particle in the medium, $\sigma_{3}(\rho)$ (here $\rho$ is the 
transverse distance between quark and gluon, the antiquark is 
located at the center of mass of the $qg$-system). 
The $\sigma_{3}(\rho)$ can be written as 
$\sigma_{3}(\rho)=C(\rho)\rho^{2}$. The factor $C(\rho)$
has a smooth (logarithmic) dependence on $\rho$ for
$\rho\ll 1/\mu_{D}$ (hereafter, $\mu_{D}$ is
the Debye screening mass). 
If one replaces $C(\rho)$ by $C(\rho_{eff})$, where $\rho_{eff}$ is the typical
value of $\rho$, the Hamiltonian takes the oscillator form.
This approximation, which  greatly simplifies the calculations, 
was employed in several analyses 
\cite{Z2,W1,OA} (it was also used in the BDMPS approach \cite{BDMPS,BDMS1}). 
However, the oscillator approximation turns out to be too crude and 
unsatisfactory. First of all,
for RHIC and LHC conditions, the dominating $\rho$ scale is not
small enough and the results depend strongly on the choice 
of $\rho_{eff}$. Another reason why the oscillator approximation 
is unsatisfactory is more serious. 
In the high energy limit the gluon formation length, $L_{f}$, becomes
larger than the quark pathlength in the QGP, and finite-size effects
become important. In this regime $\rho_{eff}\ll 1/\mu_{D}$, and 
one might naively  expect 
that the oscillator approximation should work
very well. However, one can show \cite{Z_OA} 
that in this regime the dominating $N=1$ rescattering contribution 
(and any odd rescattering)
evaluated in the oscillator approximation simply vanishes.   
For RHIC and LHC conditions, the finite-size effects play a very
important role and the oscillator approximation can lead to 
uncontrolled errors. For this reason, one has to use an accurate 
parametrization of the three-body cross section. It requires 
numerical calculations for solving the Schr\"odinger equation.

In the present paper we represent the induced
gluon spectrum in a new form which is convenient for numerical 
computations. We, for the first time, calculate the 
induced gluon emission and the nuclear modification factor 
for RHIC conditions using a realistic imaginary potential.

\vspace{.2cm}
\noindent {\bf 2.} We consider a quark with energy $E$ produced 
in a medium at $z=0$ (we chose the $z$-axis along the quark momentum). 
The induced gluon spectrum in the gluon fractional longitudinal 
momentum $x$ reads \cite{Z1}
\beq
\frac{d P}{d
x}=2\mbox{Re}\!
\int\limits_{0}^{\infty}\! d
z_{1}\!
\int\limits_{z_{1}}^{\infty}d
z_{2}
g(x)\left[{K}(z_{2},\r_{2}|z_{1},\r_{1})
-{K}_{v}(z_{2},\r_{2}|z_{1},\r_{1})\right]
\Big|_{\r_{1}=\r_{2}=0}\,.
\label{eq:10}
\eeq
Here $K$
is
the Green's
function
for the Hamiltonian (acting in the transverse
plane)
\beq
{H}=
-\frac{1}{2M(x)}\,
\left(\frac{\partial}{\partial \r}\right)^{2}
+v(\r,z) +\frac{1}{L_{f}}\,,
\label{eq:20}
\eeq
\beq
v(\r,z)=-i\frac{n(z)\sigma_{3}(\rho)}{2}\,,
\label{eq:30}
\eeq
and 
\beq
{K}_{v}(z_{2},\r_{2}|z_{1},\r_{1})=\frac{M(x)}{2\pi i(z_{2}-z_{1})}
\exp\left[\frac{iM(x)(\r_{2}-\r_{1})^{2}}{2(z_{2}-z_{1})}-
\frac{i(z_{2}-z_{1})}{L_{f}}\right]
\label{eq:40}
\eeq
is the Green's
function
for the Hamiltonian (\ref{eq:20}) with
$v(\r,z)=0$.
In
(\ref{eq:20}), the Schr\"odinger mass is
$M(x)=Ex(1-x)$,
$L_{f}={2Ex(1-x)}/{[m_{q}^{2}x^{2}+m_{g}^{2}(1-x)]}\,\,$
is the gluon formation length,
here
$m_{q}$ and $m_{g}$ are the quark and gluon  masses that play 
the role of the infrared cutoffs at $x\sim 1$
and $x\sim 0$ (in the QGP $m_{q,g}$ are given by the quark and gluon 
quasiparticle masses).
%
In (\ref{eq:30}), $n(z)$ is the number density of QGP, and $\sigma_{3}$
is
the above mentioned cross section of 
the color singlet $q\bar{q}g$ system with a particle in the medium. 
Summation over triplet (quark) and octet (gluon) color states is
implicit in (\ref{eq:30}).
The $\sigma_{3}$ may depend on $z$ (through the Debye screening mass),
however, below we will use $z$-independent $\mu_{D}$.
The vertex factor $g(x)$, entering
(\ref{eq:10}), reads 
\beq
g(x)=
\frac{\alpha_{s}P(x)}{2M^{2}(x)}\,
\frac{\partial}{\partial \r_{1}}\cdot\frac{\partial}{\partial \r_{2}}\,,
\label{eq:50}
\eeq
where $P(x)=C_{F}[1+(1-x)^{2}]/x$ is the splitting function for
the $q\rightarrow gq$ transition ($C_{F}$ is the quark Casimir factor).
Note that we neglect the spin-flip $q\rightarrow qg$ transition,
which gives a small contribution to the quark energy loss.

The three-body cross section entering the potential (\ref{eq:30}) can be
written as \cite{NZZ,Z1}
\beq
\sigma_{3}(\rho)=\frac{9}{8}
[\sigma_{q\bar{q}}(\rho)+
\sigma_{q\bar{q}}((1-x)\rho)]-
\frac{1}{8}\sigma_{q\bar{q}}(x\rho)\,,
\eeq
where
\beq
\sigma_{q\bar{q}}(\rho)=\alpha_{s}^{2}C_{T}C_{F}\int d\qb
\frac{[1-\exp(i\qb\r)]}{(q^{2}+\mu^{2}_{D})^{2}}\,
\label{eq:60}
\eeq
is the dipole cross section for the color singlet $q\bar{q}$ pair
($C_{T}$ is the color Casimir for the thermal parton 
(quark or gluon)).

The spectrum (\ref{eq:10}) can be rewritten as ($L$ is the quark pathlength
in the medium)
\beq
\frac{d P}{d
x}=
\int\limits_{0}^{L}\! d z\,
n(z)
\frac{d
\sigma_{eff}^{BH}(x,z)}{dx}\,,
\label{eq:70}
\eeq
\beq
\frac{d
\sigma_{eff}^{BH}(x,z)}{dx}=\mbox{Re}
\int\limits_{0}^{z} dz_{1}\int\limits_{z}^{\infty}dz_{2}\int d\r\,
g(x)K_{v}(z_{2},\r_{2}|z,\r)
\sigma_{3}(\rho)
K(z,\r|z_{1},\r_{1}){\Big|}_{\r_{1}=\r_{2}=0}\,.
\label{eq:80}
\eeq
$d\sigma^{BH}_{eff}/dx$ (\ref{eq:80}) can be viewed as an 
effective Bethe-Heitler
cross section, which accounts for the LPM and finite-size effects.
One can represent (\ref{eq:80}) as
\beq
\frac{d
\sigma_{eff}^{BH}(x,z)}{dx}=-\frac{\alpha_{s}P(x)}{\pi M(x)}\mbox{Im}
\int\limits_{0}^{z} d\xi
\left.\frac{\partial }{\partial \rho}
\left(\frac{F(\xi,\rho)}{\sqrt{\rho}}\right)
\right|_{\rho=0}\,\,,
\label{eq:90}
\eeq
where the function $F$ is the solution to the radial Schr\"odinger 
equation for the azimuthal quantum number $m=1$
\beq
i\frac{\partial F(\xi,\rho)}{\partial \xi}=
\left[-\frac{1}{2M(x)}\left(\frac{\partial}{\partial \rho}\right)^{2}
-i\frac{n(z-\xi)\sigma_{3}(\rho)}{2}+
\frac{4m^{2}-1}{8M(x)\rho^{2}}
+\frac{1}{L_{f}}
\right]F(\xi,\rho)\,.
\label{eq:100}
\eeq
The boundary condition for $F(\xi,\rho)$ reads
$F(\xi=0,\rho)=\sqrt{\rho}\sigma_{3}(\rho)
\epsilon K_{1}(\epsilon \rho)$, where 
$\epsilon=[m_{q}^{2}x^{2}+m_{g}^{2}(1-x)^{2}]^{1/2}$, and
$K_{1}$ is the Bessel function.
In deriving (\ref{eq:90}), we used the relations \cite{Z_YAF}
$$
\frac{\partial}{\partial \r_{1}}\cdot\frac{\partial}{\partial \r_{2}}
=\frac{1}{2}
\left[
\left(\frac{\partial}{\partial x_{1}}-i\frac{\partial}{\partial y_{1}}\right)
\cdot
\left(\frac{\partial}{\partial x_{2}}+i\frac{\partial}{\partial y_{2}}\right)
+
\left(\frac{\partial}{\partial x_{1}}+i\frac{\partial}{\partial y_{1}}\right)
\cdot
\left(\frac{\partial}{\partial x_{2}}-i\frac{\partial}{\partial y_{2}}\right)
\right]\,,
$$
$$
\left(\frac{\partial}{\partial x_{2}}\pm i\frac{\partial}{\partial y_{2}}\right)
\left.\int\limits_{z}^{\infty}dz_{2}K_{v}(z_{2},\r_{2}|z,\r)
\right|_{\r_{2}=0}=
\pm\frac{M(x)}{i\pi}\exp(\pm i\phi) K_{1}(\epsilon \rho)\,.
$$
The time variable $\xi$ in (\ref{eq:90}), in terms of the variables $z$ 
and $z_{1}$
of equation (\ref{eq:80}), is given by $\xi=z-z_{1}$; i.e.,
contrary to the Schr\"odinger equation for the Green's functions
entering (\ref{eq:10}),  (\ref{eq:90}) represents the spectrum 
through the solution
to the Schr\"odinger equation, which describes evolution of the $q\bar{q}g$
system back in time. It allows one to have a smooth boundary condition,
which is convenient for numerical calculations.

\vspace{.2cm}
\noindent {\bf 3.}
The jet quenching is 
usually characterized by the nuclear modification factor
(we consider the central rapidity region $y\sim 0$ and
suppress the explicit $y$-dependence)
\beq
R_{AA}(p_{T})=
\frac{d\sigma^{AA}(p_{T})/dydp_{T}^{2}}
{N_{bin}d\sigma^{pp}(p_{T})/dydp_{T}^{2}}\,,
\label{eq:110}
\eeq
where $d\sigma^{AA}/dydp_{T}^{2}$ and 
$d\sigma^{pp}/dydp_{T}^{2}$ are the inclusive cross section for $A+A$ and 
$p+p$ collisions, and
$N_{bin}$ is the number of the binary nucleon-nucleon collisions. 
The effect of the parton energy loss on the high-$p_{T}$ hadron
production in $A+A$ collisions can approximately be described in 
terms of effective
hard partonic cross sections, which account for the induced gluon emission
\cite{BDMS_quenching}.
Using the power-low parametrization for cross section of quark 
production in $p+p$ collisions $\propto 1/p_{T}^{n(p_{T})}$
one can obtain 
\beq
R_{AA}(p_{T})\approx P_{0}(p_{T})+\frac{1}{J(p_{T})}
\int\limits_{0}^{1} dz z^{n(p_{T})-2}
D_{q}^{h}(z,\frac{p_{T}}{z})
\int\limits_{0}^{1} dx(1-x)^{n(p_{T}/{z})-2} 
\frac{dI(x,\frac{p_{T}}{z(1-x)})}{dx}\,,
\label{eq:120}
\eeq
\beq
J(p_{T})=\int\limits_{0}^{1}dz z^{n(p_{T})-2} D_{q}^{h}(z,
\frac{p_{T}}{z})\,,
\label{eq:121}
\eeq
where $P_{0}$ is the probability of quark propagation without induced 
gluon emission,
$dI(x,p_{T})/dx$ is the probability distribution in 
the quark energy loss for a quark with $E=p_{T}$,
$D_{q}^{h}(z,p_{T}/z)$ is the quark fragmentation function.
Note that, since $n(p_{T})\gg 1$, the $z$-integrands 
in (\ref{eq:120}),~(\ref{eq:121}) are sharply 
peaked at $z\approx \bar{z}$ ($\bar{z}$ is the value of $z$
at which 
$z^{n(p_{T}-2} D_{q}^{h}(z,p_{T}/z)$ has a maximum). For this reason
(\ref{eq:120}) to quite good accuracy can be approximated as
\beq
R_{AA}(p_{T})\approx P_{0}(p_{T})+
\int\limits_{0}^{1} dx(1-x)^{n(p_{T}/\bar{z})-2} 
\frac{dI(x,\frac{p_{T}}{\bar{z}(1-x)})}{dx}\,.
\label{eq:122}
\eeq
We take the $P_{0}$ and spectrum in the radiated energy entering 
(\ref{eq:120}) in the form
\beq
P_{0}(E)=\exp\left(-\int_{x_{min}}^{1}dx 
\frac{dP(x,E)}{dx}\right)\,,
\label{eq:123}
\eeq
\beq
\frac{dI(x,E)}{dx}=\frac{dP(x,E)}{dx}\cdot\exp\left(-\int_{x}^{1}dy 
\frac{dP(y,E)}{dy}\right)
\,,
\label{eq:130}
\eeq
where $x_{min}\approx m_{g}/E$, and
it is assumed that the spectrum equals zero at $x\le x_{min}$.
Formula (\ref{eq:130}) corresponds to the leading order 
term of the series in $L/L_{rad}$ (here $L_{rad}$ is the radiation
length corresponding to the energy loss 
$\Delta E\sim E$) of the spectrum derived in 
\cite{Z_SLACSPS} for the photon emission and, strictly speaking, 
is only valid for $\Delta E\ll E$.
However, even for the more broad domain $\Delta E\lsim E$ 
(which is interesting from the point of view of 
jet quenching at 
RHIC) (\ref{eq:130}) 
reproduces the energy loss spectrum evaluated assuming independent gluon
radiation to a reasonable accuracy. 
An accurate calculation of the nuclear modification factor 
accounting for the higher order terms in $L/L_{rad}$ 
in the approximation of independent gluon emission  
\cite{BDMS_quenching} does not make sense because this
approximation itself does not have any theoretical justification.
Note that our spectrum is automatically normalized to unity.

The effective exponent $n(p_{T})$ for quark production
entering (\ref{eq:120}) is close to that for hadron production, 
$n_{h}(p_{T})$. 
A small difference between these quantities (stemming from
the $p_{T}$ dependence of the integral (\ref{eq:121}))
is given by $n(p_{T})-n_{h}(p_{T})=d\ln{J(p_{T})}/d\ln{p_{T}}$.
For $Au+Au$ collisions at $\sqrt{s}=200$ GeV
we use $n_{h}(p_{T})=np_{T}/(p_{T}+b)$ with $n=9.99$ and $b=1.219$
corresponding to the parametrization $d\sigma/dydp_{T}^{2}=A/(p_{T}+b)^{n}$
obtained in \cite{power} for $\pi^{0}$ 
production in $p+p$ collisions. The above procedure allows one 
to avoid uncertainties of the pQCD calculations  
of the partonic cross sections.

\vspace{.2cm}
\noindent {\bf 4}. 
To fix the $m_{q,g}$ and $\mu_{D}$ we use the results of the
analysis of the lattice calculations within the quasiparticle model 
\cite{LH}.
For the relevant range of temperature of the plasma $T\sim (1-3)T_{c}$ 
($T_{c}\approx 170$ MeV is the
temperature of the deconfinement phase transition) the analysis 
\cite{LH} gives for the quark and gluon quasiparticle masses
$m_{q}\approx 0.3$ and $m_{g}\approx 0.4$ GeV.
With the above value of 
$m_{g}$ from the perturbative relation $\mu_{D}=\sqrt{2}m_{g}$ one
obtains $\mu_{D}\approx 0.57$ GeV. 
To study the infrared sensitivity of our results we also perform
computations for $m_{g}=0.75$ GeV (with $\mu_{D}=0.57$ GeV). 
This value of the infrared cutoff for gluon emission in parton-nucleon
interaction has been obtained from the analysis of the 
low-$x$ proton structure function within the dipole BFKL equation
\cite{NZZ,NZ_HERA}. It seems to be reasonable for gluon emission in
the developed mixed phase and for fast gluons with $L_{f}\gsim L$.
This value agrees well with the natural infrared cutoff 
for gluon emission in the vacuum $m_{g}\sim 1/R_{c}$, where 
$R_{c}\approx 0.27$ fm is the gluon correlation radius in the QCD vacuum
\cite{Shuryak1}.   
The above two values of $m_{g}$ give reasonable 
lower and upper limits of the infrared cutoff for the induced gluon
emission for RHIC and LHC conditions. 

We perform numerical calculations for fixed and running $\alpha_{s}$.
In the first case we take $\alpha_{s}=0.5$ 
for gluon emission from light quarks and gluons, and $\alpha_{s}=0.4$
for the case of $c$-quark. For the running $\alpha_{s}$, 
we use the parametrization (with $\Lambda_{QCD}=0.3$ GeV) 
frozen at $\alpha_{s}=0.7$ at low momenta. 
This parametrization is consistent with the integral 
of $\alpha_{s}(Q)$ in the interval $0<Q<2$ GeV obtained from the analysis 
of the heavy quark energy loss (in vacuum) \cite{DKT}. Previously 
this parametrization was used to describe successfully the HERA data
on the low-$x$ proton structure function within the dipole BFKL
approach  \cite{NZZ,NZ_HERA}.
To incorporate the running $\alpha_{s}$ in
our formalism, we include $\alpha_{s}$ in the integrand on the right-hand 
side of (\ref{eq:90}) and take for virtuality $Q^{2}=aM(x)/\xi$. 
The parameter $a$
was adjusted to reproduce the $N=1$ rescattering contribution
evaluated in the ordinary diagrammatic approach \cite{Z_kinb}. It gives
$a\approx 1.85$. For the dipole cross section (\ref{eq:60}) we take
$Q^{2}=\qb^{2}$.

We assume the Bjorken \cite{Bjorken} longitudinal expansion of the QGP with
$T^{3}\tau=T_{0}^{3}\tau_{0}$ which gives $n(z)\propto 1/z$ for $z>\tau_{0}$.
We use the initial conditions suggested 
in \cite{FMS}:
$T_{0}=446$ MeV and $\tau_{0}=0.147$ fm for RHIC, and
$T_{0}=897$ MeV and $\tau_{0}=0.073$ fm for LHC.
For RHIC, the above condition were obtained from 
the charged particle pseudorapidity density 
$dN/dy\approx 1260$ measured by the PHOBOS experiment
\cite{PHOBOS} in $Au+Au$ collisions at $\sqrt{s}=200$ GeV
assuming an isentropic expansion and rapid thermolization at 
$\tau_{0}\sim 1/3T_{0}$.
The LHC parameters correspond to $dN/dy\approx 5625$ at $\sqrt{s}=5.5$ TeV,
which was estimated in 
\cite{KMS}. The above initial conditions for RHIC (translated into 
$\tau_{0}=0.6$ fm) agree with those used in successful hydrodynamic description
of $Au+Au$ collisions at RHIC \cite{HeinzKolb}.
Note that, since the dominating $\rho$-scale in (\ref{eq:80})
$\propto \sqrt{z}$ for $z\ll L_{f}$,
our results are not very sensitive to $\tau_{0}$ (for a given entropy).
The maximum parton pathlength in the hot QCD medium is restricted by
the life-time of the QGP (and mixed) phase, $\tau_{max}$. 
\footnote[1]{For our choice of the initial conditions the life-time of QGP is 
$\sim 3$ fm for RHIC. However,  
in the interval $\tau\sim 3-6$ fm the density of the
mixed phase is practically the same as that for the pure QGP phase.}. 
We take $\tau_{max}\sim R_{A} \sim 6$ fm.
This seems to be a reasonable value for central heavy-ion collisions,
since,  
due to the transverse expansion, the hot QCD matter
should cool quickly at $\tau\gsim R_{A}$ \cite{Bjorken}.

In Fig.~1 we show the induced gluon spectrum for 
$q\rightarrow gq$
transition for RHIC conditions for the quark pathlength 
$L=6$ fm obtained
with $m_{q}=0.3$ and $m_{q}=1.5$ GeV for $m_{g}=0.4$ and $m_{g}=0.75$ GeV.
In Fig.~1 we also show the Bethe-Heitler spectrum (dashed line).
One sees that the LPM and finite-size effects strongly  suppress  
the gluon emission. The gluon emission from $c$-quark is suppressed
in comparison with light quark due to larger mass, which leads
to decreasing of the dominating $\rho$ scale (note that
the spectrum is not sensitive to the light quark mass, 
except for $x\sim 1$). One can see from Fig.~1
that, although the Bethe-Heitler spectrum differs strongly for two values of
$m_{g}$, the difference becomes relatively small for 
the spectrum, which accounts for the LPM and finite-size effects.
It is connected with the fact that, due to the multiple scattering and
finite-size effects, the dominating $1/\rho$-scale becomes larger 
than $m_{g}$; namely this in-medium scale plays the role of the infrared 
cutoff at high energies \cite{Z2} (however, of course, for not very 
high $p_{T}$ the value of $m_{g}$ is still important). 
We do not show the spectra for 
running $\alpha_{s}$.
They are similar in form (but  somewhat suppressed at moderate 
fractional momenta). 
The LPM suppression for LHC is considerably stronger than for RHIC, but 
the spectra are similar in form, and we do
not show them as well.

In Fig.~2 we plot the quark energy loss
$
\Delta E=E\int_{x_{min}}^{1} dx x{dP}/{dx}
$
evaluated for  fixed (solid line) and running (dashed line) $\alpha_{s}$ 
for RHIC and LHC conditions for $L=6$ fm with $m_{g}=0.4$ GeV (thick lines)
and $m_{g}=0.75$ GeV (thin lines).
The results for $\alpha_{s}=0.5$ agree 
roughly with that for running $\alpha_{s}$ for $E\lsim 10$ GeV
but at higher energies the energy dependence is steeper for fixed
$\alpha_{s}$. This says that the typical $\rho$-scale becomes
smaller with increasing energy. It is also seen from the decrease
of the relative difference between the curves for $m_{g}=0.4$ and $0.75$ GeV.

In Fig.~3a we compare the nuclear modification factor (\ref{eq:120})
for $T_{0}=446$ MeV 
calculated using the NLO KKP fragmentation functions \cite{KKP}
for running $\alpha_{s}$ with that obtained at 
RHIC \cite{RHIC_data} for central
$Au+Au$ collisions at $\sqrt{s}=200$ GeV. 
For illustration of the dependence on $T_{0}$, in Fig.~3b we also present  
the results for $T_{0}=375$ MeV.
The theoretical curves were obtained
for $L=4.9$ fm. It is the typical parton pathlength in
the QGP (and mixed) phase for $\tau_{max}=6$ fm. We present the
results for $m_{g}=0.4$ and $0.75$ GeV. For $p_{T}\lsim 15$ the results for 
$\alpha_{s}=0.5$ are close to that for running $\alpha_{s}$ 
and we do not plot them. 
The results for the quark (solid line) 
and gluon (dashed line) jets are shown separately
(note that for $\sqrt{s}=200$ GeV the quark and gluon contributions 
are comparable). The suppression is somewhat stronger for gluon jets.
One can see from Fig.~3a that the theoretical $R_{AA}$ for 
$m_{g}=0.4$ is in reasonable
agreement with the experimental one.
One should bear in mind, however, that our calculations neglect the 
collisional energy loss \cite{Bjorken2}.
For $p_{T}\sim 5-15$ GeV the collisional energy loss may increase 
the total energy loss by $\sim 30-40$\%. In this case (if one takes
the initial conditions \cite{FMS}) the value $m_{g}=0.75$ GeV would be 
more preferable for agreement with the RHIC data. 
As mentioned previously, this value is reasonable
for the mixed phase and for gluons with $L_{f}\gsim L$. 
Since, for $\sqrt{s}=200$ GeV, the medium spends about half
of its time in the mixed phase, the effective infrared cutoff may be
larger than the gluon quasiparticle mass in the QGP.
For this reason, the collisional energy loss may be included without
using an unrealistic infrared cutoff for the induced energy loss.
The possible remaining small disagreement with the data may be avoided 
by taking a somewhat smaller value of $T_{0}$
(or $\alpha_{s}$). In any case, it is clear that, for such a complicated 
phenomenon, it is hardly possible to expect a perfect agreement with
experiment and the agreement found in the present paper is surprisingly
good. 

The above estimate for the collisional energy 
loss has been obtained for the pQCD plasma.
Presently, there are some indications \cite{Shuryak2} that the medium 
produced at RHIC may be a strongly coupled QGP. 
The radiative energy loss should not be very sensitive to the dynamics of 
the QGP (for the same number density of the QGP).
However, it may be important for the collisional
energy loss. Unfortunately, the corresponding calculations have not 
been made yet.
It is interesting that our results give support for the scenario with
strongly coupled QGP. Indeed, this scenario requires 
$\alpha_{s}\gsim 0.5$ \cite{Shuryak2} for thermal partons.
The $R_{AA}$ is sensitive to the radiation of soft gluons with an 
energy of about a few $\mu_{D}$.
One can expect that, for such gluons, $\alpha_{s}$ should be close to that
for thermal partons. We obtained agreement with the data with 
$\alpha_{s}$, which is frozen at a value of $0.7$ at low momenta.
If $\alpha_{s}$ is frozen at a value below 0.4-0.5, the theoretical 
$R_{AA}$ strongly disagrees
with that observed at RHIC.

\vspace{.2cm}
\noindent {\bf 5}. 
In summary, we have represented, within the LCPI approach \cite{Z1}, 
the induced gluon spectrum in a new form convenient for numerical 
calculations
and carried out computations of the induced 
gluon emission from fast partons in the expanding QGP for 
RHIC and LHC conditions. The calculations
for, the first time, have been performed with a realistic parametrization
of the dipole cross section. 
The theoretical nuclear modification factor calculated for the initial 
conditions obtained from the charged particle rapidity density
observed at RHIC \cite{FMS} and the hydrodynamic simulation of the 
RHIC data \cite{HeinzKolb}
is in a reasonable agreement with that observed at RHIC.

\bigskip
\noindent {\large \bf Acknowledgements.} 
I thank R.~Baier and D.~Schildknecht for discussions and their kind 
hospitality at the University
of Bielefeld, where this work was completed.
I am also grateful to the High Energy Group of 
the ICTP for their   kind hospitality during my
visit to Trieste, where part of this work was done.

\newpage

\begin{center}
{\Large \bf Figures}
\end{center}
\begin{figure}[h]
\begin{center}
\epsfig{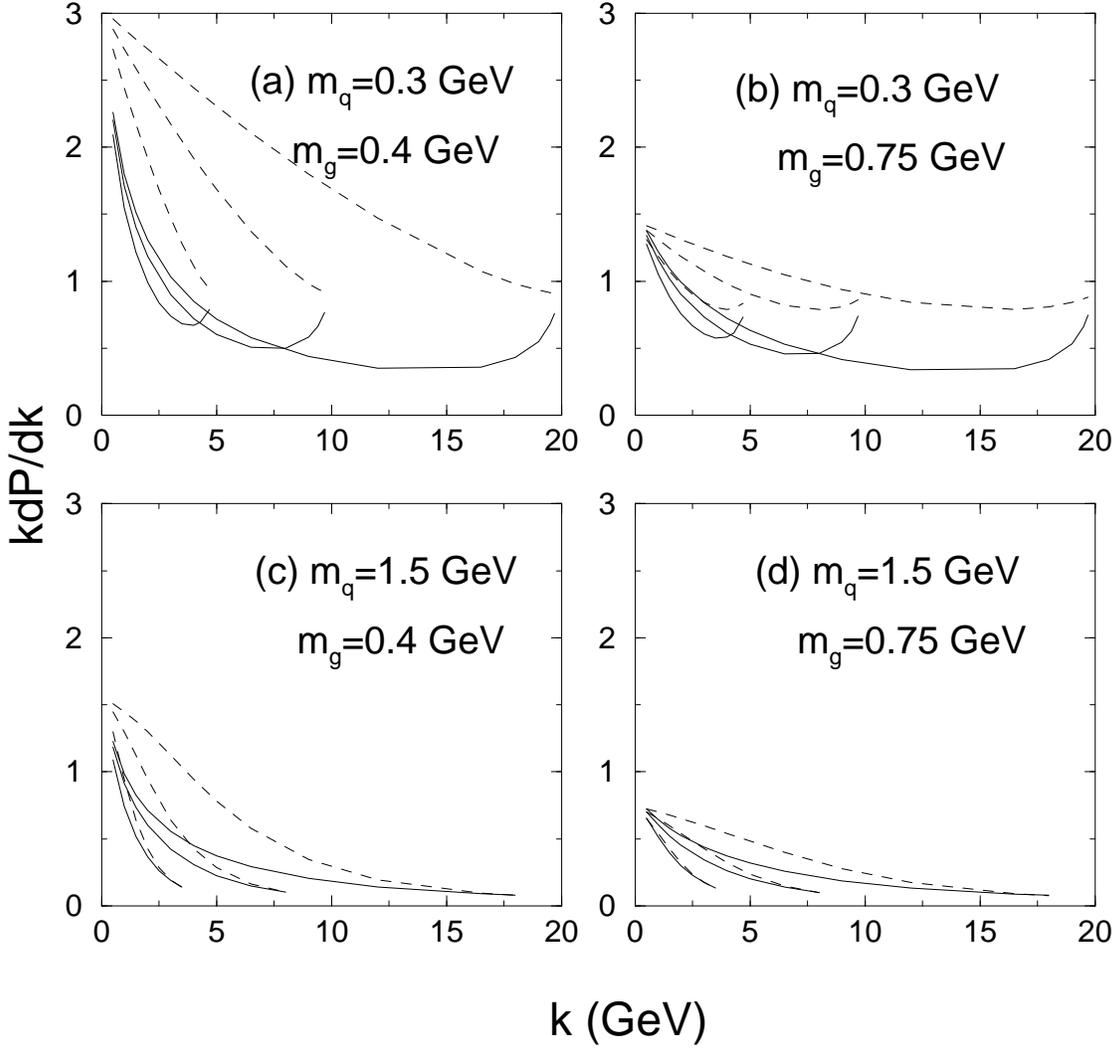}
\end{center}
\caption[.]{
The induced gluon spectrum (solid line) for $q\rightarrow gq$ transition 
versus the gluon momentum $k$ for
RHIC conditions for $E=5,$ 10 and 20 GeV, $L=6$ fm obtained 
using (\ref{eq:70}),~(\ref{eq:90}) with $\alpha_{s}=0.5$ for 
$m_{q}=0.3$ GeV (a,b) and $m_{q}=1.5$ GeV (c,d);
$m_{g}=0.4$ GeV (left) and 0.75 GeV (right).
The Bethe-Heitler spectrum is shown by the dashed line.
}
\end{figure}

\begin{figure}[h]
\epsfig{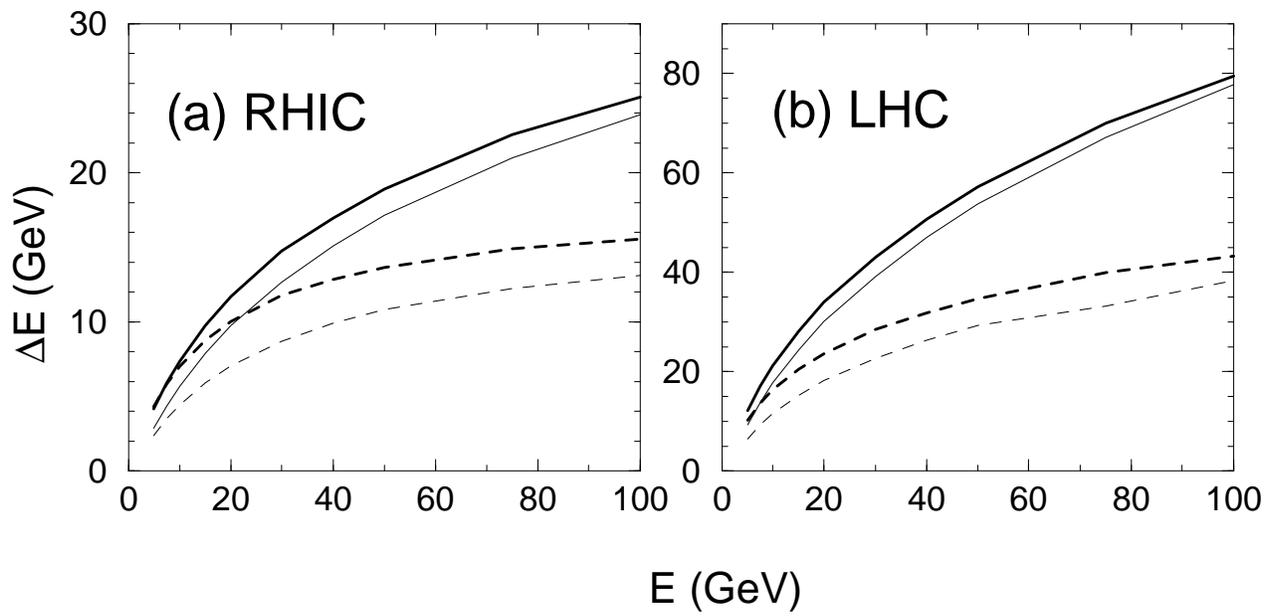}
\caption[.]{
The energy dependence of the quark energy loss 
 for RHIC (a) and LHC (b)
for $L=6$ fm obtained with  
$\alpha_{s}=0.5$ (solid line) and running $\alpha_{s}$ (dashed line),
$m_{g}=0.4$ GeV (thick lines) and $m_{g}=0.75$ GeV (thin lines),
$m_{q}=0.3$ GeV.
}
\end{figure}

\begin{figure}[t]
\begin{center}
\epsfig{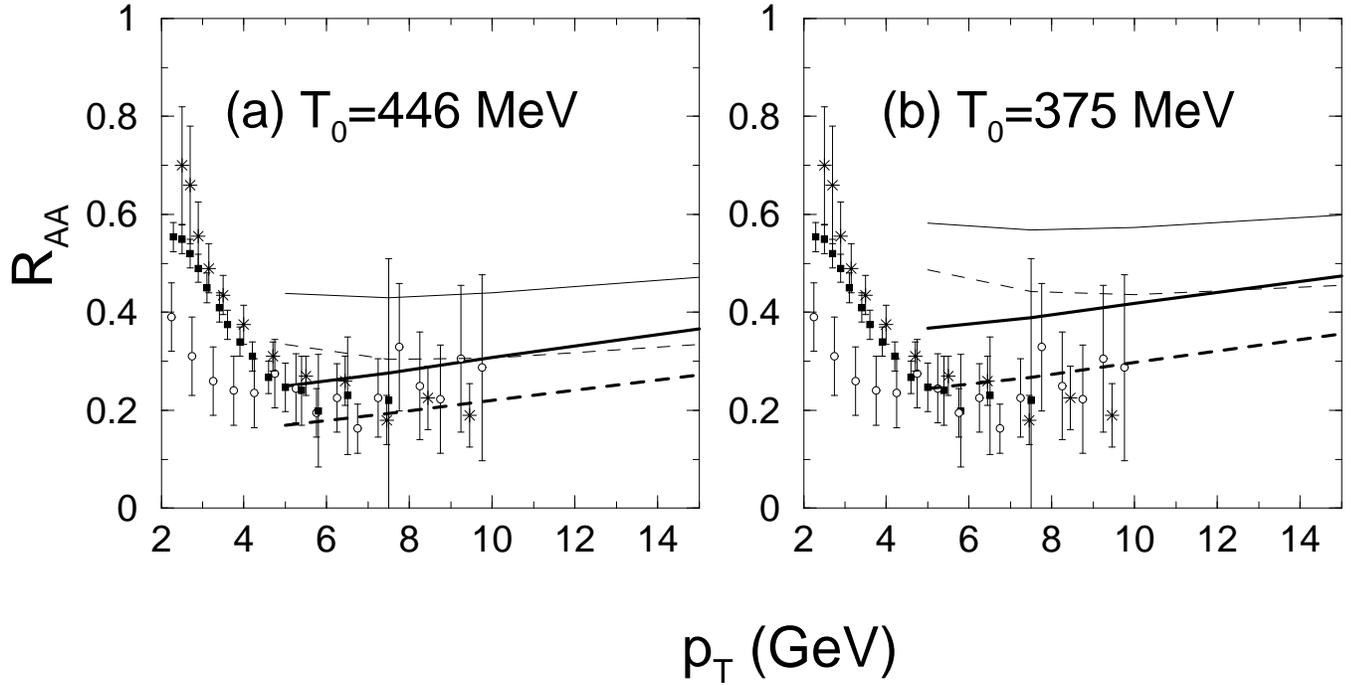}
\end{center}
\caption[.]{
The nuclear modification factor (\ref{eq:120})
for central $Au+Au$ collisions at $\sqrt{s}=200$ GeV
for quark (solid line) and 
gluon (dashed line)
jets obtained with $m_{g}=0.4$ GeV (thick lines) and $m_{g}=0.75$ GeV 
(thin lines) for running $\alpha_{s}$.
The experimental points (from \cite{RHIC_data}) are for the following:
circle - $Au+Au\rightarrow \pi^{0}+X$ (0-10\% central) 
[PHENIX Collaboration], 
square - $Au+Au\rightarrow h^{\pm}+X$ (0-10\% central)
 [PHENIX Collaboration],
star - $Au+Au\rightarrow h^{\pm}+X$ (0-5\% central)
[STAR Collaboration].
}
\end{figure}

\end{document}